\documentclass[preprintnumbers,amsmath,amssymb,twocolumn,showpacs]{revtex4}
\usepackage{graphicx}
\usepackage{dcolumn}
\usepackage{bm}

\begin{document}

\title{Faraday rotation of a tightly focussed beam from a single trapped atom}

\author{G. H\'etet$^{1}$, L. Slodi\v{c}ka$^1$, N. R\"ock and R. Blatt$^{1,2}$}

\affiliation{
$^1$ Institute for Experimental Physics, University of Innsbruck, A-6020 Innsbruck, Austria \\
$^2$ Institute for Quantum Optics and Quantum Information of the
Austrian Academy of Sciences, A-6020 Innsbruck, Austria}

\begin{abstract}
Faraday rotation of a laser field induced by a single atom is demonstrated by tightly focussing a linearly polarized laser beam onto a laser-cooled ion held in a harmonic Paul trap.
The polarization rotation signal is further used to measure the phase-shift associated with electromagnetically-induced-transparency and to demonstrate read-out of the internal state on the qubit transition with a detection fidelity of 98 $\pm$ 1\%. These results have direct implications
for single atom magnetometery and dispersive read-out of atomic superpositions.
\end{abstract}
\maketitle

Coherent manipulation of phase-shifts imposed by atoms onto light fields has been a subject of investigations for decades. Besides the classical optics applications, the subject gained interest with the perspective of coherent control of quantum states of light and matter. Atomic phase shifts allow for instance localization of quantum states of light within quantum memories \cite{Lvo09}, dispersive read-out of internal \cite{Vol10,Kos10,Was10} and external \cite{Rabl} atomic states and can be used to create Schr\"odinger cat states \cite{Sav90}. Investigating the detailed properties of phase shifts with well-localized atoms and single-photons would furthermore yield several direct applications in quantum information science \cite{San11, Ris11}.
\begin{figure}[ht!]
\centerline{\scalebox{0.3}{\includegraphics{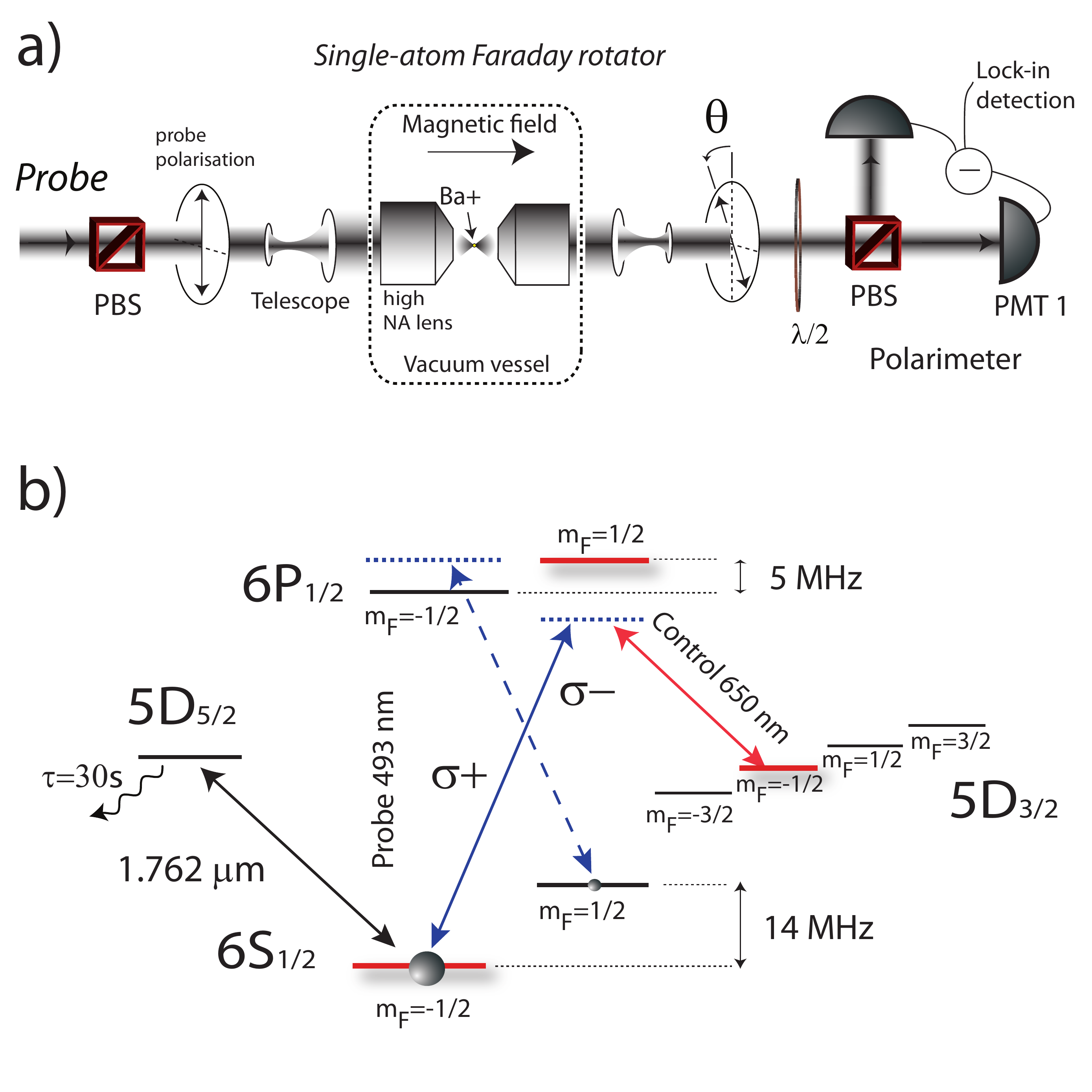}}}
\caption{a) Scheme of the single atom Faraday rotator. The probe field is prepared in a horizontally polarized mode after passing through a polarizing beam-splitter (PBS). The laser field is then expanded by a telescope and coupled through a high numerical aperture lens in vacuum (NA=0.4). The single ion rotates the polarization of the light field, which after re-collimation is detected by polarimetry.
b) Level scheme of $^{138}$Ba$^+$ and laser fields used in the experiment. The input probe at 493 nm is decomposed in the circular basis and excites the two branches of the spin-half system with different detunings.
A laser field at 650 nm is used  for both repumping population from the $D_{3/2}$ level and as a control field in the EIT phase shift measurements. A laser at 1.762 $\mu m$ can also be tuned to excite the quadrupolar transition from the $S_{1/2}$ to the $D_{5/2}$ level.
 }\label{setup}
\end{figure}
To achieve this, common approaches use optical cavities coupled to a few atoms \cite{Kam10, Mue10} or the coupling of a single light mode to an optically thick column of atoms \cite{Fle05}.
A recent research area also investigates direct coupling of light to single atoms in free space using high numerical aperture elements. Recently, single cold rubidium atoms \cite{Tey08}, single cold molecules \cite{Wri08, Pot}, quantum dots \cite{Vam07} have been successfully employed to observe effects that were often thought to be specific to high finesse cavities or to atomic ensembles. A one degree phase shift induced by a single Rubidium atom onto a freely propagating probe was for example measured in \cite{Alj09}.

Single ions are also good candidates for further fundamental investigations of such single-pass light-atom couplings \cite{Shu,Streed,Maiwald,Jechow}. 
The strong confinement offered by Paul traps and a high numerical aperture system allowed us recently to observe electromagnetically induced transparency (EIT) from a Barium ion \cite{slo10} as well as its operation as a mirror of a Fabry-P\'erot-like cavity \cite{Het11}.
In this letter, we demonstrate Faraday rotation of a coherent laser field induced by a single trapped ion.
This is achieved by tightly focussing a weak detuned linearly polarized probe field onto a single Barium ion and fine tuning of the ground state atomic populations of a four-level atomic system. We then implement this phase measurements for reading out the state of the single atom on the optical quadrupole transition with a fidelity of 98\%. We furthermore use the Faraday rotation signal as a tool to observe the steep phase shift across the electromagnetically induced transparency spectrum and we discuss the implication of our experimental results for quantum information processing.

Faraday rotation is a ubiquitous phenomenon in optics. It takes place when
an optically active material rotates the polarization direction of a linearly polarized input field in the presence of a magnetic field. In circularly birefringent materials, such as atomic gases \cite{Bud02,Lab01,Nob10,Ter09}, the two circular components of the field acquire different phase shifts, the strengths of which depend on the magnetic field amplitude. 

Figure \ref{setup}-a) shows the experimental set-up that we use for measuring the Faraday rotation with a single atom. A Barium ion is trapped and cooled in a ring Paul trap and a probe field is tightly focussed onto it using a telescope followed by a high-numerical aperture objective in-vacuum \cite{Esc01, Het11, slo10}. A magnetic field of 5 Gauss is applied along the probe field direction and also used here to define the quantization axis. A linearly polarized probe field is shone on the ion and its polarization is analyzed using a polarimetric set-up and photo-multiplier tubes (PMT). In practice, we record the signal using only one detector and a waveplate alternatively tuned at 45 or -45 degrees with respect to the polarizing beam splitter axis.

The level scheme of our Barium ion is shown Fig. \ref{setup}-b). The probe field is tuned to the $S_{1/2}\rightarrow P_{1/2}$ transition and with our choice of quantization axis, its polarization is decomposed onto left and right circularly polarized modes.
The two polarization modes are detuned differently from their transitions and may experience different indices of refraction. We set the intensity of the probe field well below saturation so that elastic scattering always dominates. To provide cooling of the ion, we also use a red detuned 493~nm laser field that is perpendicular to the probe direction and a laser at 650 nm, co-propagating with the cooling beam, for pumping out population from the $D_{3/2}$ level. To precisely measure the polarization rotation signal, we use a
locking method where the repumper is switched at a rate of 5 kHz. The PMT signal is then demodulated and low-pass filtered with a time constant of 1s.

The intensity of the light at PMT$_1$ can be written as
\begin{eqnarray}
I_{45}=\frac{1}{2}|E^+_{\rm out} e^{i\pi/4}+E^-_{\rm out}  e^{-i\pi/4} |^2,
\end{eqnarray}
where $E^+_{\rm out} =(1-2\epsilon \mathcal{L}^+)E_{\rm in}$ and $E^-_{\rm out} =(1-2\epsilon \mathcal{L}^-)E_{\rm in}$.  The real and imaginary parts of $\mathcal{L}^{\pm}=\rho_\pm\gamma/(\gamma+i\Delta^{\pm})$ correspond to absorption and phase-lag of the two scattered circularly polarized field modes with regards to the input field, respectively. $\Delta^{\pm}=\Delta \pm \Omega_B$ are the detunings of the $\sigma^+$ and $\sigma^-$ polarized fields from their respective transitions. $\Delta$ is the probe laser detuning, defined as the detuning from the $S_{1/2}\rightarrow P_{1/2}$ transitions if they were not Zeeman shifted and $\Omega_B$ is the Zeeman splitting. $\rho_\pm$ are the populations in the $S_{\pm1/2}$ ground states, respectively.
\begin{figure}[ht!]
\centerline{\scalebox{0.32}{\includegraphics{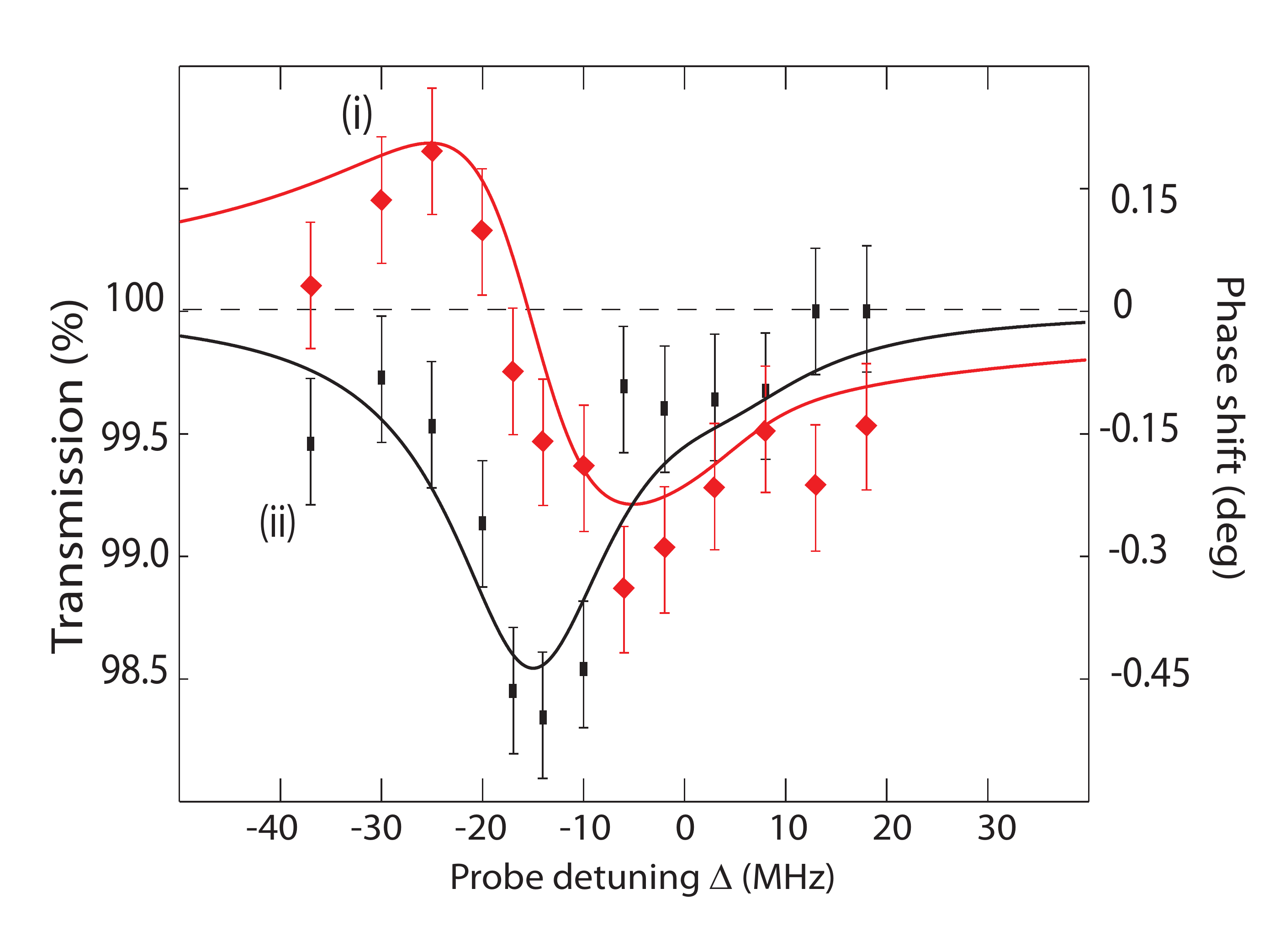}}}
\caption{a) Trace (i)-(ii) show the respective Faraday rotation angle $\theta$ (or phase shift) and the transmission $s_0$ of the probe laser through the single ion as a function of the laser detuning. The two solid lines show a fit using four-level equations.
Error bars correspond to one standard deviation estimated from the Poissonian noise within our detection bandwidth.
}\label{FLA}
\end{figure}
The $\pm\pi/4$ phase shifts are due to the rotation of the polarization direction induced by the $\lambda/2$ plate on the probe. The resulting overall $\pi/2$ phase shift between the two circularly polarized modes allows access to the imaginary part of one of the field amplitudes. To measure the rotation of the polarization angle, we record the total transmission $s_0$
 without the $\lambda/2$ waveplate. After inserting the waveplate, we can access the $s_2$ Stokes parameter \cite{BornWolf:1999:Book} defined as $s_2=2 I_{45}-s_0$. For our small extinction values the other Stokes parameters $s_1 \approx s_0$. The Faraday rotation angle $\theta\approx 1/2 \arctan(s_2/s_0)$ is directly related to the phase shift induced by the atom. It can indeed be shown that, using the approximation $\arg(1-2\epsilon z)\approx -2\epsilon {\rm Im}(z)$ in the limit of small $\epsilon$
\begin{eqnarray}\label{ps}
\theta=\frac{1}{2}\arg\big[1-2\epsilon ( \mathcal{L}^+ + \mathcal{L}^- )\big],
\end{eqnarray}
which is the total phase lag experienced by the output field with respect to the input.
A measurement of $I_{45}$ and $s_{0}$ thus provides a measurement of the Faraday rotation of the light across the atom together with the phase difference acquired by the two circularly polarized modes.

$I_{45}$ and $s_{0}$ are first measured as a function of the probe field detuning from the $S_{1/2}$ to $P_{1/2}$ level. For these measurements, the cooling beam was kept on and tuned to one of the two dark resonances that provides efficient pumping to the $S_{1/2}(m_F=-1/2)$ level whilst still allowing cooling of the ion at the Doppler limit. The levels that are involved in this so called ``dark-state pumping" are marked in red in Fig.\ref{setup}-b). Fig.~\ref{FLA} shows the results of the measurement of $\theta$ (trace (i)) and $s_{0}$ (trace (ii)) as a function of the probe frequency difference from the $S_{1/2}$ to $P_{1/2}$ level.
As can be seen from the measurement of $s_0$, the dark-state pumping causes a strong unbalancing between the two ground states populations. This is manifest in the 1.5\% extinction that is seen -14 MHz red-detuned from the central line and in the almost completely suppressed extinction for the other mode at 5 MHz.
With this dark-state preparation technique, trace (i) displays a clear dispersive profile across the resonance of the $\Delta m=+1$ transition and the circularly polarized mode $\sigma^-$ is almost not phase shifted. Even with our small magnetic field, the pumping technique thus allows us to isolate a single two-level atom and to reach a maximum of 0.3 degrees phase-shift.
Solid lines show the result of a fit of the data using the above four-levels calculations, with $\epsilon=0.8\%$, $\Omega_B=9$~MHz, $\rho_-=0.9$ and $\rho_+=0.1$. With these parameters, good agreement is found with the experimental results. 

Compared with measurements of phase shifts using Mach-Zehnder interferometers (MZI), our polarization rotation method is a rather simple and precise way of inferring atomic phase shifts.
In fact, the technique may still be seen as an interference process between the two polarization states $\sigma^+$ and $\sigma^-$, which follow the same optical path and then interfere after the waveplate and polarizing beam-splitter. 
Besides providing immediate spatial mode matching of the two polarization modes, another advantage is that any thermal or acoustic noise that might lower MZI signals do not matter here since they are common to both circularly polarized fields.

\begin{figure}[t]
\centerline{\scalebox{0.45}{\includegraphics{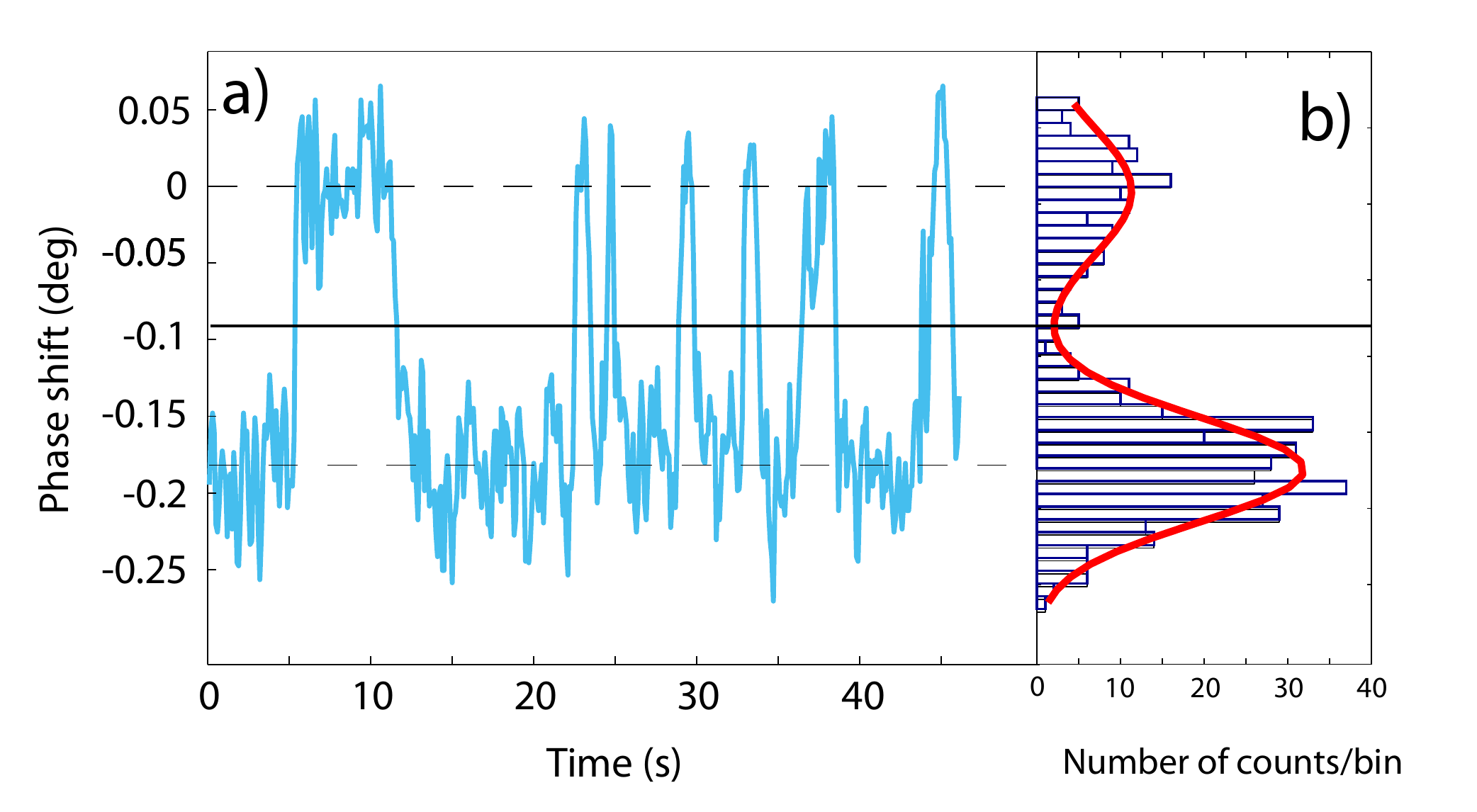}}}
\caption{a) Polarimeter signal as a function of time when the 1.76$\mu m$ laser field is tuned to the $S_{1/2}\rightarrow D_{5/2}$ quadrupolar transition. An integration time of 300 ms was used here. Several quantum jumps are observed. b) Histogram showing the probability of measuring a given phase shift over these 45 seconds of measurements within a 0.01 bin-width. }\label{qjumps}
\end{figure}

We now show that the polarization rotation detection method also provides a means of measuring atomic qubits. With a laser at 1.762 $\mu m$ we are able to perform coherent manipulations on the $S_{1/2}$ to $D_{5/2}$ transitions \cite{Slo12}. Here, by weakly driving this quadrupole transition, and with a 300 ms phase shift read-out,  we can detect changes in the phase-shift much faster than the decay time of our Barium qubit ($\approx$ 30 seconds).
Fig. \ref{qjumps}-a) shows the evolution of the phase-shift $\theta$ during 45 seconds when a very weak $1.76 \mu m$ laser excitation is tuned close to the quadrupole transition. The probe frequency is here set to $\Delta=0$ with a detected count rate of of 200 kcounts/s. We observe that the phase shift is interrupted by sudden quantum jumps which reveal that population is transferred to the $D_{5/2}$ level, thereby completely canceling the polarization rotation.
Fig. \ref{qjumps}-b) shows a histogram where phase shifts values are counted within a 0.01 degree bin-width during the 45 seconds-long measurement. Two distinct bell-shaped distributions corresponding to two mean coherent state amplitudes are seen. The probability of inferring the correct atomic state is $1/2(p_{S}+p_{D})=98\pm 1\%$ when using the thick solid line as a threshold. $p_{S,D}$ are the probabilities of finding the atom in the $S_{1/2}$ or $D_{5/2}$ level, which are estimated by fitting the histograms by a sum of two Gaussian distributions and by calculating the fraction of population that lies inside the chosen threshold. 

At present, the measurement time needed to resolve the atomic state is larger than when using fluorescence detection in the same set-up \cite{Slo12}, but the method may prove useful for efficient read-out of atomic states in systems where such an electron shelving is more involved.
Such a dispersive read-out of the atomic population has furthermore potential for non-destructive, or quantum non-demolition (QND), measurements of atomic superposition.
To estimate the regimes where the QND condition may be fulfilled in our setup, we performed a Ramsey sequence on the $S_{1/2}\rightarrow D_{5/2}$ transition with a $50\mu s$ pulse of coherent probe light at 493 nm between the two $\pi/2$ pulses. The highest mean photon number that we can use in the laser pulse before spontaneous emission reduces the Ramsey contrast was found to be $\overline{n}=40\pm10$. This limits the maximum probe intensity that can be used for estimating atomic superpositions before state projection occurs. Increasing the numerical aperture of our lens may allow sufficient signal-to-noise ratio for such QND measurements to be performed and will be investigated more thoroughly in the future.
Our polarization rotation observations can also find applications for real-time read-out of small magnetic fields. Using, for instance, a radio-frequency field driving the two ground states together with a detection of the polarization rotation at the Larmor frequency, would enable active feedback to the magnetic coils for compensating magnetic field noise \cite{Bud07} at the exact location of the single atom.


\begin{figure}[t]
\centerline{\scalebox{0.3}{\includegraphics{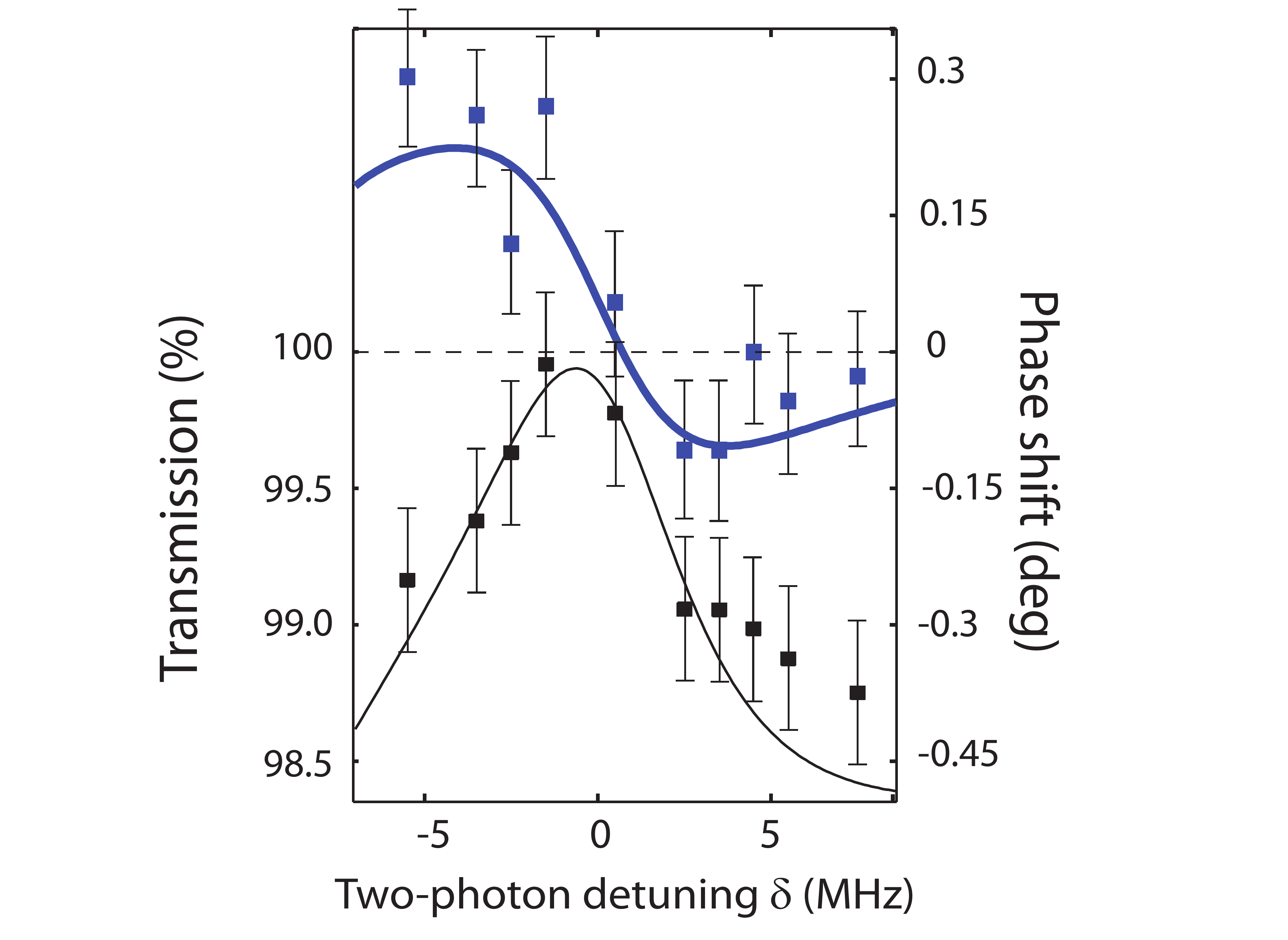}}}
\caption{Electromagnetically-induced-transparency transmission and phase shifts from the single atom as a function of the probe laser detuning (red and black points). The solid lines show a fit using 8-level Bloch equations. }\label{EIT}
\end{figure}

Moreover, the Faraday rotation technique allows us to investigate
the dependence of the single atom induced phase shift in the presence of quantum interference effects.
In the above measurements, the cooling and repumping beams were tuned to a dark resonance in order to provide prepumping to one of the $S_{1/2}$ levels. 
As in \cite{slo10}, we now turn off the transverse cooling beam and cool with the linearly polarized probe field itself. 
In such a configuration the probe undergoes electromagnetically-induced-transparency (EIT) \cite{slo10} where the population in the excited state of the $\Lambda$ scheme (see Fig. \ref{setup}-b)) is canceled due to a quantum interference between the two excitation pathways leading to the $P_{1/2}$ excited state.
Fig.~\ref{EIT}- trace (i) shows the result of the measurement of the probe transmission versus the two-photon detuning. In this EIT regime, a rapid change of the transmission is found as a function of the two-photon detuning $\delta$ and an almost complete cancelation of the transmission is measured at $\delta=0$.
Associated with such a steep change of the probe transmission, we also expect a fast roll-off of the phase.
Fig.~\ref{EIT}-trace (ii) shows the measurement of $\theta$, using the same polarimetric technique as for the previous four-level scheme measurements. Here again, close to the dark resonance, the Faraday rotation angle yields the phase-shift induced by the atom. The clear dispersive shape of $\theta$ across the two-photon resonance is here a sign of the EIT induced phase-shift from the ion where a maximum phase lag of 0.3 degrees is observed. The solid lines show a fit to the experimental results using 8-level Bloch equations where we replace the two-level atom Lorentzian functions $\mathcal{L}^{\pm}$ in Eq.~\ref{ps} by newly found susceptibilities. The theory describes well the data with the repumping and probe field intensities as the only two free parameters. 
The asymmetry of the dispersion and transmission profiles that we measured is due to
a slight overlap with neighboring dark-resonances and our detuned driving of the $\Lambda$ scheme. 
The distinctive feature of this interference effect is that the flipping of the phase shift sign occurs only over a couple of MHz. Increasing the slope steepness further can in fact be done by performing the experiment with smaller probe and repumping powers which can be implemented by appropriate switching of the laser cooling beams involved the experiment. Achieving a very steep phase shift dependence across the atomic spectrum would open the way for reading out the motional \cite{Rabl} and internal \cite{Wil03} energy of the atom.

In conclusion, we demonstrated polarisation rotation of a coherent laser field by a single well localized ion. Tightly focussing a weak detuned linearly polarized probe field onto a single Barium ion and further tuning of the atomic population was then shown to provide an effective means of measuring single atom phase shifts. We could then observe the steep phase change across the electromagnetically-induced-transparency spectrum. Phase shifts of 0.3 degrees were measured, limited mostly by the finite numerical aperture of the employed detection optics. Last, we use the phase measurements for high fidelity read-out of the single atom state by inducing quantum jumps with a narrow laser beam tuned to a long lived optical transition. Besides demonstrating further the potential of free-space coupling to single ions for fundamental quantum optics and quantum information science, these experimental results will trigger interest for quantum feedback to the motional state of single atoms, as proposed in \cite{Rabl} using EIT, for dispersive read out of atomic qubits and for ultra-sensitive single atom magnetometery \cite{Bud02,Bud07}.

This work has been partially supported by the Austrian
Science Fund FWF (SFB FoQuS), by the European Union
(ERC advanced grant CRYTERION), by the Institut fur
Quanteninformation GmbH and by the FWF SINFONIA grant. 
G. H. acknowledges support
by a Marie Curie Intra-European Action of the European
Union.


\end{document}